\documentclass[twocolumn]{webofc}

\usepackage{siunitx}
\usepackage[varg]{txfonts}   

\usepackage{hyperref}
\usepackage{url}
\usepackage{graphicx}        
\usepackage{dcolumn}         
\usepackage{bm}              
\usepackage{xcolor,colortbl} 

\usepackage{array}
\usepackage{acronym}
\usepackage{listings}
\usepackage{longtable}
\usepackage{alltt}
\usepackage[]{subfig}
\usepackage{float} 

\hypersetup{
    colorlinks=true,
    citecolor=violet,
    urlcolor=blue,
    linkcolor=purple
}

\begin{document}
\title{Recent radiological studies at NSLS-II using the Monte Carlo code \textrm{FLUKA}}
%
%

\author{\firstname{Ricardo} \lastname{dos Santos Augusto}\inst{1}\fnsep\thanks{e-mail: \url{rdossanto@bnl.gov}} 
  \and
        \firstname{Robert J.} \lastname{Lee}\inst{2}
}

\institute{Radiological Control Division, Brookhaven National Laboratory, Upton, NY 11973‐5000
\and
           National Synchrotron Light Source II, Brookhaven National Laboratory, Upton, NY 11973‐5000
          }

\abstract{ \it This work details some of the recent radiological studies performed at National Synchrotron Light Source II (NSLS-II) using the Monte Carlo (MC) particle transport code FLUKA. In particular, it assesses the radiation protection challenges posed by the anticipated NSLS-II facility upgrade, taking into consideration various upgrade options. Also, it evaluates the  impact related to the use of different shielding materials and the employment of ambient dose H* as the external radiation exposure  estimator of effective dose. Lastly, a demonstration of synchrotron radiation transport is provided for an in-silico beamline, all the way from the radiation source to end-station, utilizing FLUKA's recent X-ray reflectivity capabilities.
}
\maketitle
\section{Introduction}
\label{intro}

The NSLS-II facility is a third-generation light source with a 3 GeV and 500 mA electron beam, currently supplying synchrotron radiation to almost 30 beamlines, with future development allowing for twice that capacity.

Shielding of an NSLS-II beamline is designed to protect against \emph{Gas Bremsstrahlung} (GB) and \emph{Synchrotron Radiation} (SR), as well as any secondary radiation generated. Those are chronic sources of radiation and the main contributors to dose rates outside of the shielding in normal operating conditions. Another potential source of secondary radiation derives from the NSLS-II's Top-Off mode of operation, which implies frequent injection of electrons into the storage ring allowing for the continuous operation of the accelerator and beamlines by maintaining stable stored beam current and bunch pattern \cite{fliller}.

At NSLS-II chronic ambient equivalent dose rates are to be kept as low as reasonably achievable (ALARA) and lower than 5 \si{\micro\sievert\per\hour} outside of the accelerator enclosure during regular operation.
The shielding design goal is to achieve less than 0.5 \si{\micro\sievert\per\hour} on contact with the experimental enclosures. This implies that annual dose to staff and users is kept lower than 1 \si{\milli\sievert} assuming an  yearly occupancy of 2000 hours.

Exposure to personnel due to a fault event shall not exceed 200 \si{\micro\sievert} in radiologically unposted areas or 1 \si{\milli\sievert} in a radiation controlled area. 
Shielding is designed so that the total dose to an individual, integrated over the duration of the fault, does not exceed 1 \si{\milli\sievert}.

As per the NSLS-II Shielding Policy, interlocking ARMs are deployed in radiologically controlled areas where radiation fields can exceed 1 \si{\milli\sievert\per\hour} and their high-alarm is set to 50 \si{\micro\sievert\per\hour}
. Detector placement is optimized to promptly detect fault events and terminate operations few seconds after reaching the alarm level, thus mitigating the duration of unacceptable radiation conditions in potentially occupied areas.

\section{NSLS-II upgrade scenarios}

Radiological limits will be kept unchanged for the upcoming NSLS-II upgrade (NSLS-IIU). Possible upgrade scenarios include increments in energy, changes in intensity, or a combination of both (see table \ref{upgrades}).

\begin{table*}[htbp]
\caption{\label{upgrades}%
\it Upgrade scenarios and their parameters.   
}
\centering
\begin{tabular}{l|c|c|c|c}
\textrm{Upgrade}&
\textrm{None}&
\textrm{NSLS-IIU 3}&
\textrm{NSLS-IIU 3.5} &
\textrm{NSLS-IIU 4}\\\hline
 Electron beam energy [\si{\giga\electronvolt}] & 3 & 3 & 3.5 & 4 \\
Intensity [\si{\milli\ampere}] & 500 & 600 & 500 & 400 \\

\end{tabular}

\end{table*}

The radiological impact associated with each of the proposed upgrade variants will be quantified, as these may result in more severe injection fault events as well as harsher GB and SR source terms.

\section{The DIM beamline}

\begin{figure*}[h]
\centering
\frame{\includegraphics[scale=.4]{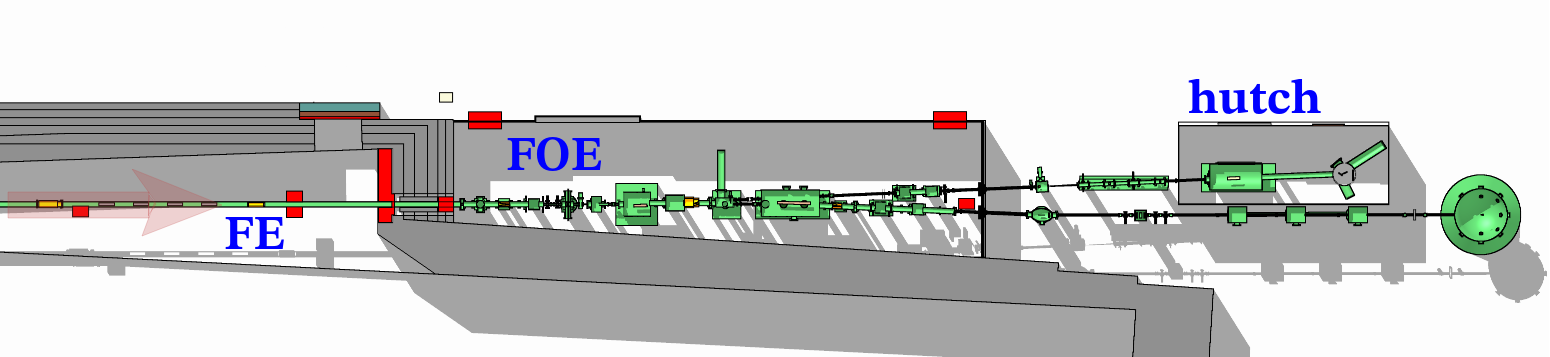}}
\caption{\it \label{DIM} Top view of the DIM beamline model as seen in flair GeoViewer \cite{flair}. The arrow indicates the beam direction.}
\end{figure*}

For full flexibility in the shielding performance calculations, a virtual beamline model was prepared specifically for these studies rather than utilizing an existing NSLS-II beamline. This \textit{in silico} beamline, denominated DIM, is depicted in figure \ref{DIM}.

Its SR source is a bending magnet directing X-rays to one of two thick Si scattering mirrors and then towards the respective end-stations via several mirrors. Both the front end (FE) and first optical enclosure (FOE) shielding thicknesses and materials were kept consistent with the typical shielding configuration for beamlines at NSLS-II. However, some characteristics of other NSLS-II beamlines (e.g., largest apertures)  were adopted in order to reproduce more conservative scenarios in the DIM calculations.

For the most conservative upgrade scenario, the GB source is also assessed with a variant of the FOE shielding featuring a lateral wall and roof shielded with steel instead of lead. Further details are listed in table \ref{shishi}

\begin{table}[htbp]
\caption{\label{shishi}%
\it  Details of main beamline shielding components, including the alternative roof and outboard wall steel shielding thicknesses which are enclosed in parentheses.
}
\begin{tabular}{l|c|c}
\textrm{Description}&
\textrm{Material}& 
\textrm{Thickness [\si{\centi\meter}]}\\
\hline
 Near FE door & Concrete & 60$^\star$\\
 FOE roof & Lead  (Steel) & 0.6 (1.5)\\
 FOE outboard wall & Lead  (Steel) & 1.8 (4.5) \\
 FOE downstream wall & Lead & 5\\

\end{tabular}
\vspace{0.1cm}

 \footnotesize $^\star$The indicated concrete thickness corresponds to its thinnest layer.
\end{table}

The thicknesses of the beam transport pipe in table \ref{shishi} are not optimized for shielding performance.

\section{Dosimetry analysis}

The tools-of-choice for shielding performance calculations at NSLS-II are the analytical code STAC8 \cite{Asano} 
 (for SR calculations only) and the Monte Carlo particle transport and interaction code FLUKA \cite{fluka1,fluka4}. Dosimetry evaluations are usually detailed in Ambient Dose Equivalent [H*(10)]. However, in 2020, ICRU95 changed the operational quantity from Ambient Dose Equivalent to Ambient Dose (H*). Since 2023, FLUKA versions employ Ambient Dose conversion coefficients by default.

Despite the same units and similar names, H* is a completely different
quantity with respect to H*(10).
For instance, the former is not measurable but rather computable if one can measure the radiation field, by folding the measured radiation field with the ICRP recommended conversion coefficients.

In the framework of NSLS-II radiation protection activities it is obviously deemed more practical to use H*(10) since it is a measurable quantity. However, anticipating the eventual adoption of the new ICRU recommendation by regulatory entities, this study features a direct comparison between H* and H*(10) to evaluate the potential radiological implications for NSLS-II activities.

\section{Radiation source terms analysed}

\subsection{\label{TOSS}Injection faults}

\begin{figure}[htbp]
\frame{\includegraphics[scale=.29]{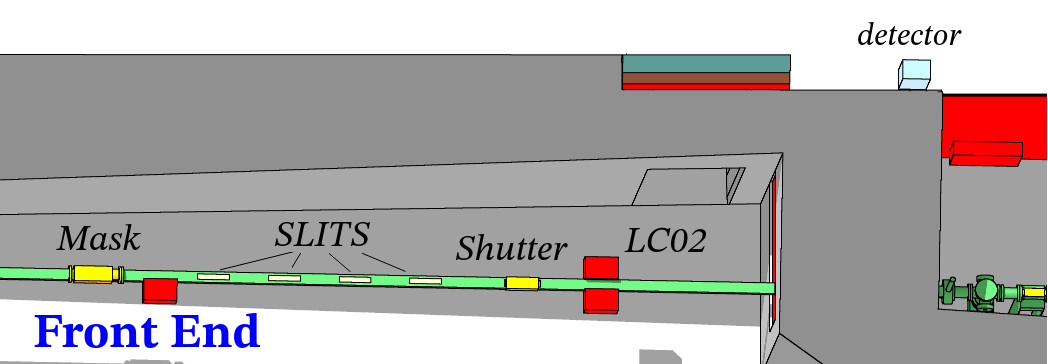}}
\caption{\it \label{fe1} Detail view of the Front End identifying the main elements of interest for the top-off injection fault analyses. }
\end{figure}

 To establish the safety of top-off injection, in the event of a faulty combination of lattice magnetic field settings and beam energy all errant particles must be captured before reaching the second lead collimator (FE's LCO2 in figure \ref{fe1}) and no injected electrons can travel downstream into the FOE nor strike the LCO2 within 5 \si{\milli\metre} of the edge of its aperture. This safe point will be maintained regardless of the upgraded electron beam variant.

 FLUKA simulations were performed to calculate the radiation levels in the potentially occupied areas on the experimental floor due to injected electron beam missteering incidents. The electron beam energy in these incidents ranged from 3 to 4 \si{\giga\electronvolt} as per table \ref{upgrades} and booster to storage ring injection charge rate was conservatively considered at 45 \si{\nano\coulomb\per\minute}.

 The incidents simulated assumed that a pencil electron beam originated at the Mask's  downstream apertures was directed to:

 \begin{enumerate}
     \item an endpoint located at the LCO2 collimator upstream face, 2 \si{\milli\metre} inboard from its aperture, interacting with various slits along the beam path.
     \item the LCO2 collimator upstream face, 2 \si{\milli\metre} outboard from its aperture,  intercepting the downstream flange of the FE Photon Shutter approximately 40  \si{\centi\metre} upstream of the LCO2 upstream aperture with slits removed.
 \end{enumerate}

These scenarios were selected based on their potentially conservative radiological impact towards the FOE (case 1) and  FE door area (case 2).

\subsection{\label{GB}Gas Bremsstrahlung}

As the electron beam traverses the straight section of the storage ring, GB is produced due to interactions with the residual gas. When the FE safety shutters and photon shutter are open,  GB will proceed into the FOE, generating secondary radiation as it interacts with the various beamline components.

 For consistency, GB spectra were modeled with FLUKA at the end of a 15.5 \si{\meter} straight section in every scenario with the same residual gas composition \cite{arisxn}.

\begin{figure}[htbp]
\includegraphics[scale=.225]{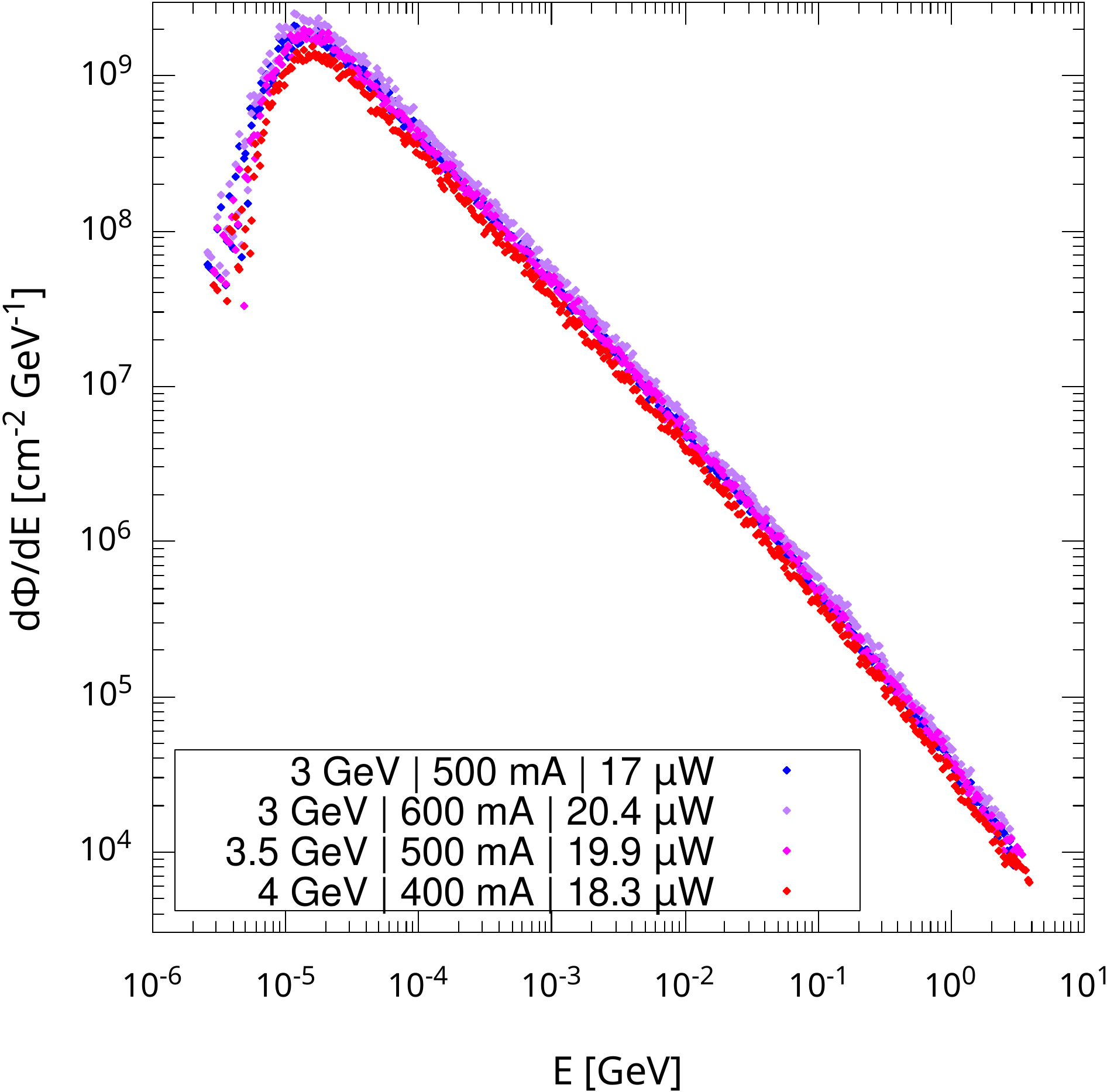}
\caption{\it \label{spectrum} Differential photon fluence, integrated over the solid angle and scored at the end of a 15.5 \si{\metre} straight section after traversing the residual gas for each scenario. }
\end{figure}

  The current NSLS-II GB maximum power of 17 \si{\micro\watt} was employed as a baseline for the second step normalization in the various upgrade scenarios.
  A comparison between source terms is provided in figure \ref{spectrum}.

The shielding performance analysis assumed that the full GB power impinged onto two scattering mirrors oriented at 1.25\si{\degree}, in one of the following configurations:

\begin{enumerate}
    \item[A.] Upstream mirror - outboard;
    \item[B.] Downstream mirror - inboard.    
\end{enumerate}

\begin{figure}[htbp]
\frame{\includegraphics[scale=.21]{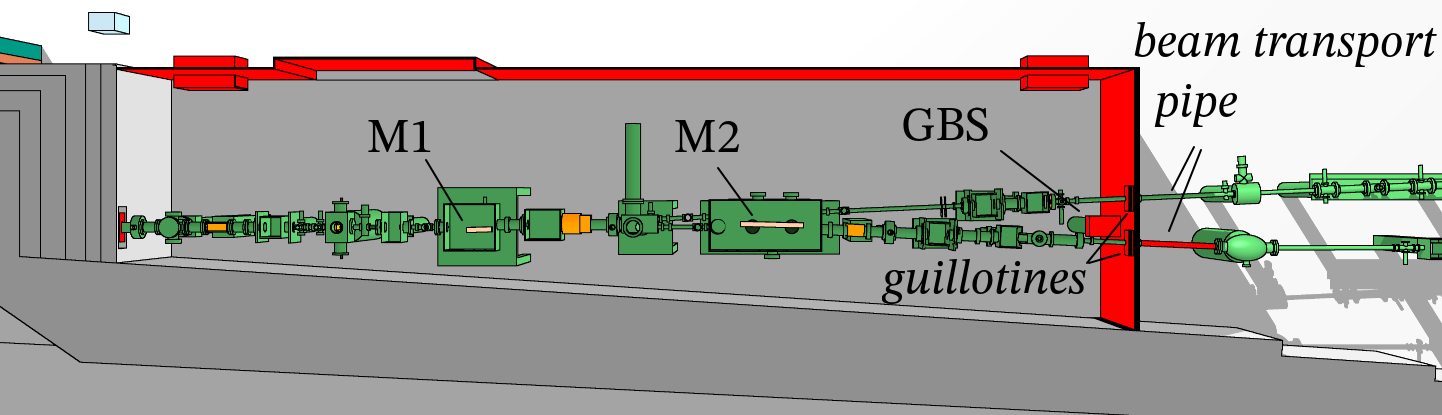}}
\caption{\it \label{mirr} Detail view of the FOE and relevant shielding elements for the GB radiological calculation. }
\end{figure}

The mirrors and the relevant components for the GB shielding calculations are denoted in figure \ref{mirr}

To confirm the radiological impact to the FOE shielding, dosimetry data were scored on contact with the outside of the FOE's roof, outboard and downstream wall.

\begin{figure}[htbp]
\centering
\frame{\includegraphics[scale=.45]{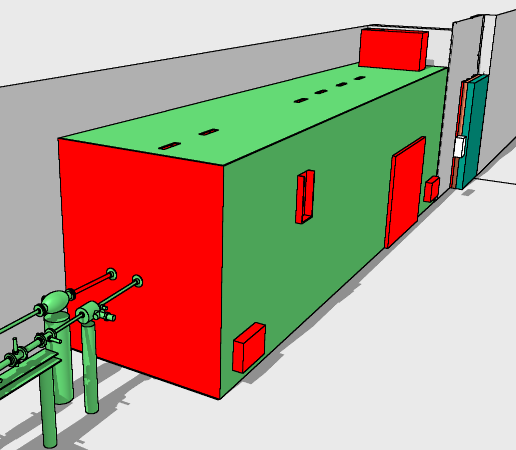}}
\caption{\it \label{stahl} FOE featuring an alternative shielding version with steel outboard walls and roof shielding (green). }
\end{figure}

This radiological analysis included also calculations performed with an FOE shielding variant featuring an outboard wall and roof made out of steel, which is illustrated in figure \ref{stahl}. Shielding details can be found  in table \ref{shishi}.

\subsection{\label{SR}Synchrotron Radiation}

A recent FLUKA  upgrade implements the reproduction of X-ray reflectivity. Among other applications, this development enables the  evaluation of the SR radiological impact in an MC framework \cite{fluka4}, complementing the information so far only provided by analytical codes at NSLS-II.%

\begin{figure}[htbp]
\frame{\includegraphics[scale=.15]{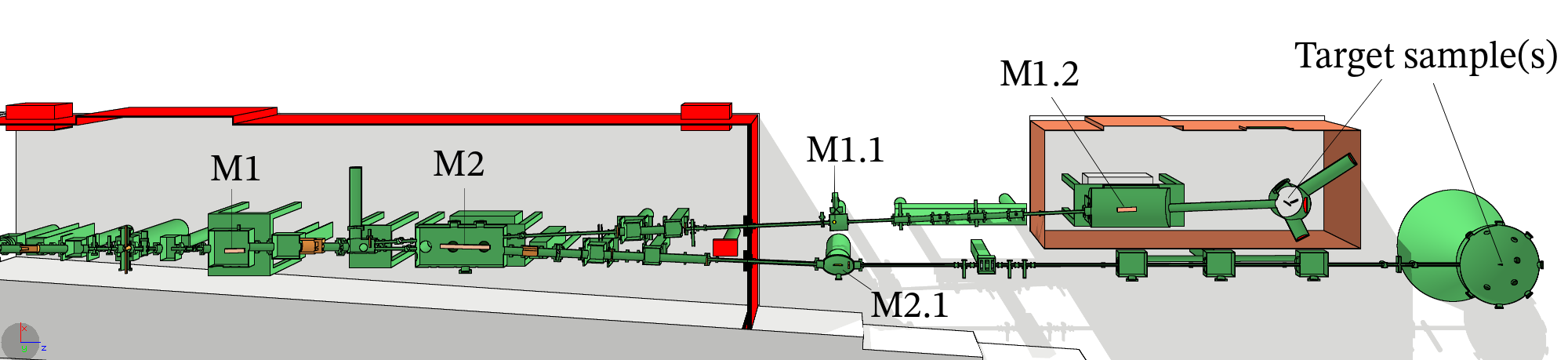}}\\

\vspace{0.1cm}

\frame{\includegraphics[scale=.25]{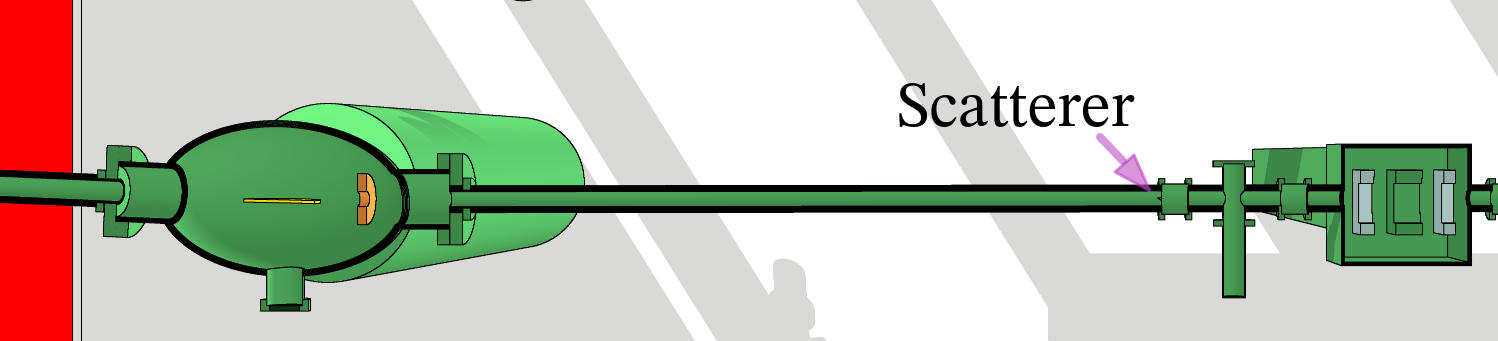}}
\caption{\it \label{mirror} Above - Plan view of the DIM beamline indicating the main optical elements. Below - The conical scatterer location downstream of M2.1. }
\end{figure}

In the DIM beamline, SR is assumed to be generated by a bending magnet and then directed to two scattering mirrors, either to M1 (upstream) or to M2 (downstream). From there, the X-rays are directed to their respective samples at the outboard endstations via M1.1 and M1.2 or inboard through M2.1 (see figure \ref{mirror}).

For a radiological impact  demonstration it will be considered that the beam is directed inboard. M2 consists of a Si mirror at an angle of 1.25\si{\degree}, whereas M2.1 is a Au mirror at 1.225\si{\degree}. A Si conical scattering element was introduced in the inboard section of the beamline, approximately 1.5 \si{\meter} downstream of M2.1 and just upstream of a thin steel bellow $\sim .455$ \si{\milli\meter} thick.

Finally, a result  combining both the  SR and GB contributions will be provided, depicting more realistically a maximum credible incident's ambient dose rates. 
 
\begin{figure}[htbp]
\scriptsize\hspace{7cm} $\Phi$ [A.~U.]\normalsize\\
\includegraphics[width=.475\textwidth]{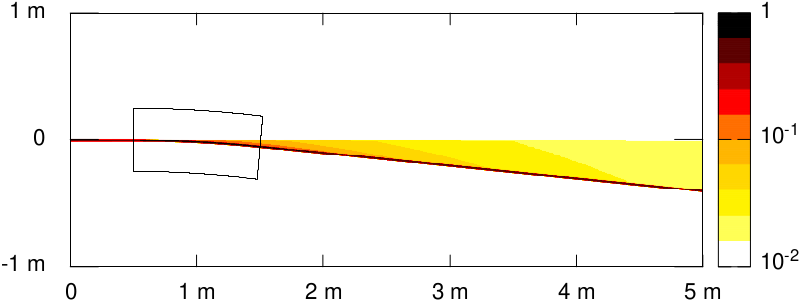}
\caption{\it \label{fig:syncrad} Distribution of SR photon fluence (in arbitrary units)  produced when a 3 \si{\giga\electronvolt} electron beam traverses a magnetic field generated by a bending magnet in the region highlighted. A second result, denoting the electron fluence corresponding to the primary beam subjected to the same magnetic field in the same region, was superimposed for visualization purposes.}

\end{figure}

\begin{figure}[htbp]
\centering
\includegraphics[scale=.45]{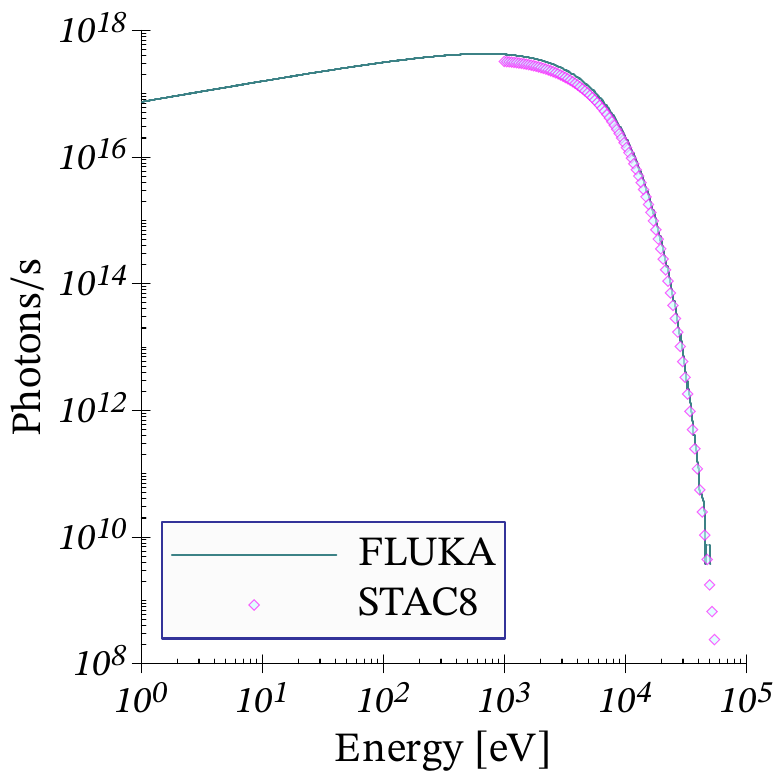}
\caption{\it \label{srs} SR source spectra generated for the 3 GeV electron beam via STAC8 (points) and the FLUKA data  (lines). }
\end{figure}

Reflectivity coefficients obtained with FLUKA were compared with both  STAC8 and LBL's X-ray data \cite{lbl}. The generated SR photons' distribution in FLUKA can be visualized qualitatively in figure \ref{fig:syncrad} and a comparison of the STAC8 and FLUKA outputs is displayed in  figure \ref{srs}.

\section{Results}

\begin{figure*}[htbp]
\centering
\hspace{14cm}  [mrem/h]\footnotesize \\

\includegraphics[scale=.28]{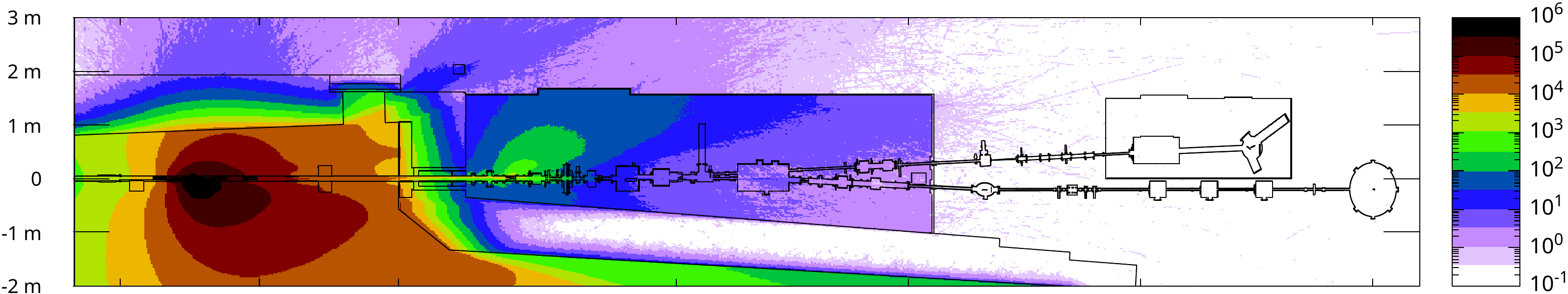} \\ 

\includegraphics[scale=.28]{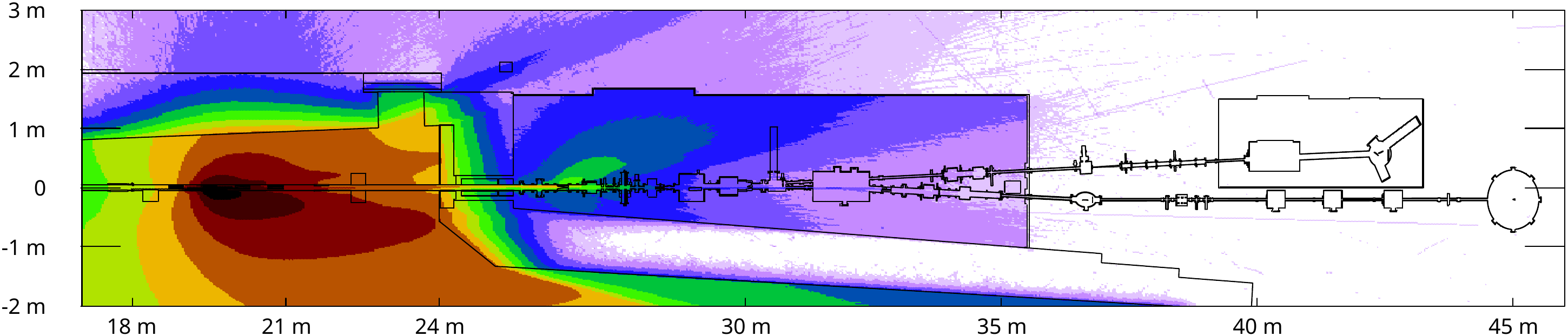}\hspace{1.35cm}

\caption{\it  \label{comp} Plan views showing the H*(10) and H* [\si{\milli rem \per\hour}] distributions (above and below, respectively) resulting from NSLS-IIU 4; case 1. Results are averaged over $Y=\pm1$ \si{\meter}, centered at the straight height. Note that $1$ \si{\milli rem} $\equiv 10$ \si{\micro\sievert}. }
\end{figure*}

The injection fault H* and H*(10) rates scored outside the shielding are reported in table \ref{tab:fantastictoss}%
. All results scored were within a $10\%$  error range. A more  qualitative result can be visualized in figure \ref{comp} too.

\begin{table}[htbp]
\caption{\label{tab:fantastictoss} Dosimetry analysis outboard of the shielding during fault incidents with electron beam loss in the FE at a 45 \si{\nano\coulomb\per\minute} injection rate. Results are given as maximum contact values of  H*(10) $|$ H* [\si{\milli\sievert\per\hour}]}
\centering
\begin{tabular}{llll}

\small

 & \small FE wall & \small Ratchet wall & \small FOE  \\ \hline
\bf \small NSLS-II(U 3) & &  &  \\

  \small Case 1   & \it \small .89 | .63 & \small \it .99  | .87 & \small \it  .57 | .92  \\ 
 \small Case 2   & \small \it .87 | .86 & \small \it 12.44 | 10.05 & \small \it .41 | .60  \\

\hline

\bf \small  NSLS-IIU 3.5  & &  &    \\

 \small Case 1   & \it \small .94 | .61 & \small \it 1.80 | 1.51 & \small \it 1.03 | .90   \\ 
 \small Case 2   & \small \it 1.06 | .85  & \small \it 13.96 | 11.46 & \small \it .64 | .62 \\

\hline

\bf \small  NSLS-IIU 4  &  & &     \\

 \small Case 1   & \small \it 1.30 | .65 & \small \it 1.37 | 1.00 & \small \it 1.08 | .72 \\ 
 \small Case 2   & \small \it 1.18 | 1.05  & \small \it 16.24 | 13.32 & \small \it .65 | .61 \\

\end{tabular}
\end{table}

\begin{figure*}[htbp]
\centering
\includegraphics[scale=0.275]{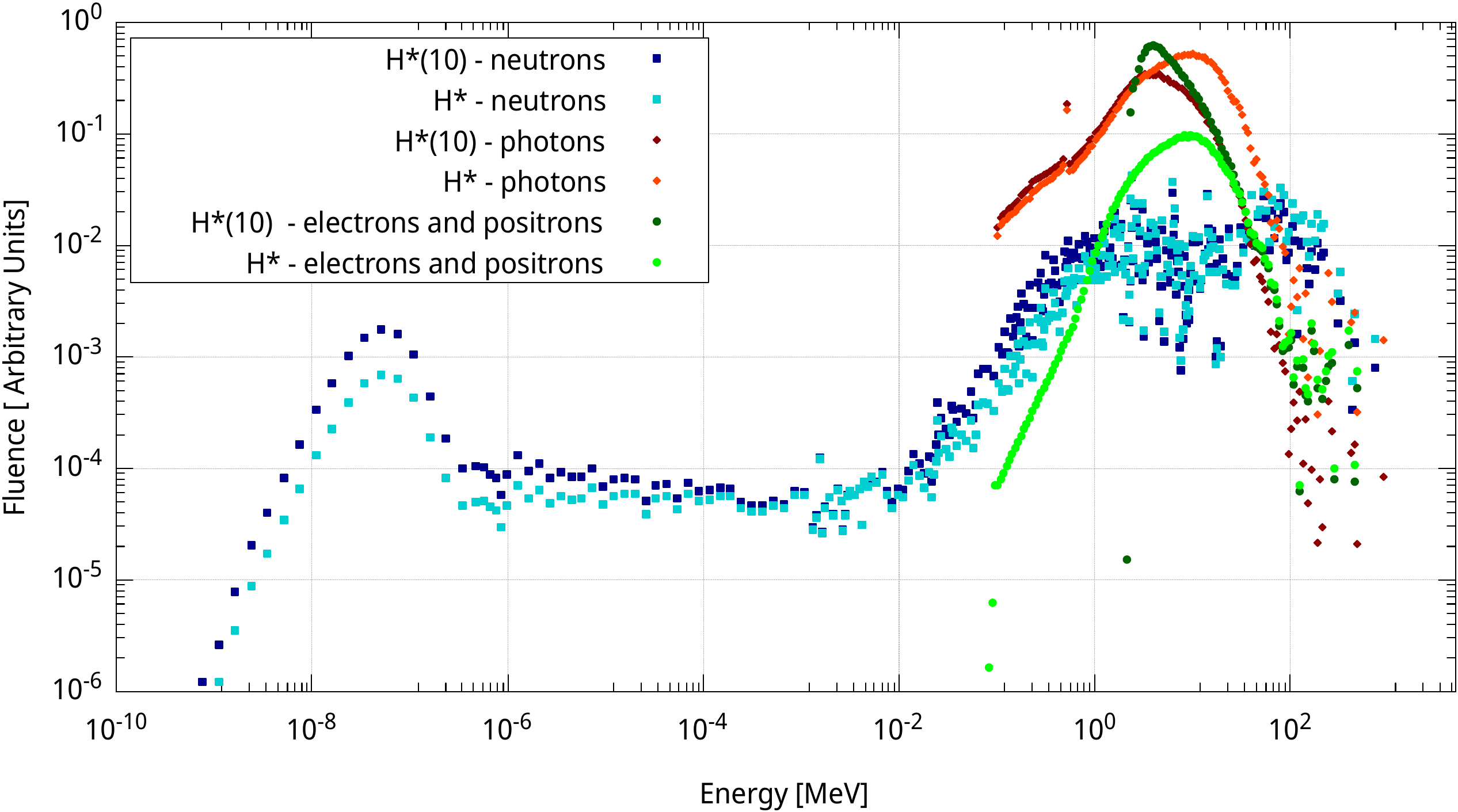}
\caption{\it  \label{blau} H*(10) and H* track-length spectra as a function of energy for different radiation types (in arbitrary units) scored at the detector in scenario NSLS-IIU 4, case 2. }
\end{figure*}

A detector, shown in the outboard side of the shielding at the interface between the FE and FOE in figure \ref{fe1}, was employed to score the radiation traversing the shielding near its weaker point. Both the H*(10) and H* track-length spectra for the different particles were scored and are depicted in figure \ref{blau}. 

Table \ref{tab:sys3} details the maximum H* and H*(10) scored on contact with the outside of the shielding for GB configurations A and B. Results are within a $10\%$  error range.%

\begin{table}[htbp]
\caption{\label{tab:sys3} Maximum overall H* | H*(10) [\si{\micro\sievert\per\hour}], on contact with the outside of the FOE shielding for the configurations A and B. }
\centering
\begin{tabular}{llll}

\bf \small  &  \small  outboard  &  \small roof  &   \small  downstream  \\ \hline
\bf \small NSLS-II &  & &    \\

    A & \it \small .74 | .68 & \it \small .57 | .75 &  \it \small .65 | .87
 \\

     B & \it \small .56 | .47
& \it \small .50 | .66 

& \it \small .54 | .65
 \\

\hline

\bf \small NSLS-IIU 3  & &   &    \\

    A & \it \small  .89 | .82
    
&  \it \small .68 | .90

&  \it \small .78 | 1.04
 \\

     B &  \it \small .67 | .56
&  \it \small .60 | .79 

&  \it \small .65 | .78
\\

\hline

\bf \small NSLS-IIU 3.5  & &   &    \\

    A &   \it \small .48 | .71
    
&  \it \small .50 | .53

&  \it \small .70 | .97
\\

     B &  \it \small .48 | .51
&  \it \small .52 | .53

&  \it \small .58 | .81
 \\

\hline

\bf \small NSLS-IIU 4  & &   &    \\

    A &  \it \small .37 | .42
    
&  \it \small .41 | .45 

&  \it \small .72 | .91
 \\

     B &  \it \small .47 | .61
&  \it \small .50 | .53

&  \it \small .52 | .72
 \\

\end{tabular}
\end{table}

\begin{figure*}[htbp]
\centering
\emph{neutron H*} \\
\vspace{0.1cm}
\includegraphics[scale=0.182]{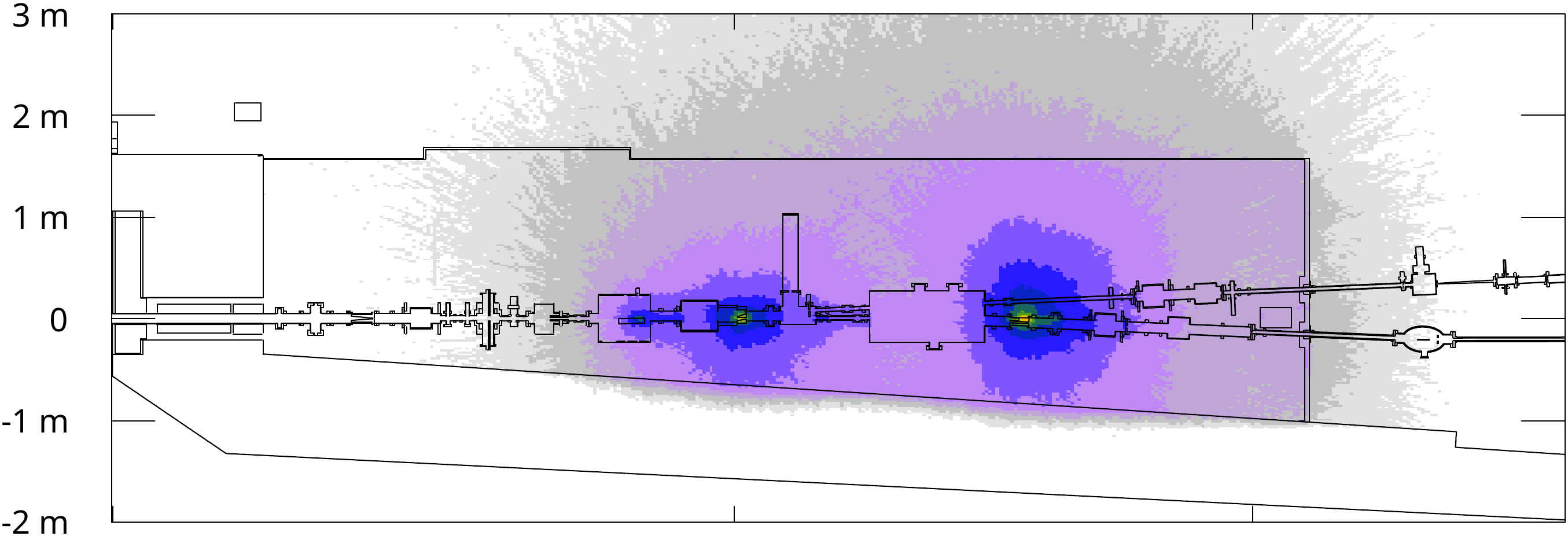}\hspace{0.2cm}\raisebox{0.6mm}{\includegraphics[scale=0.182]{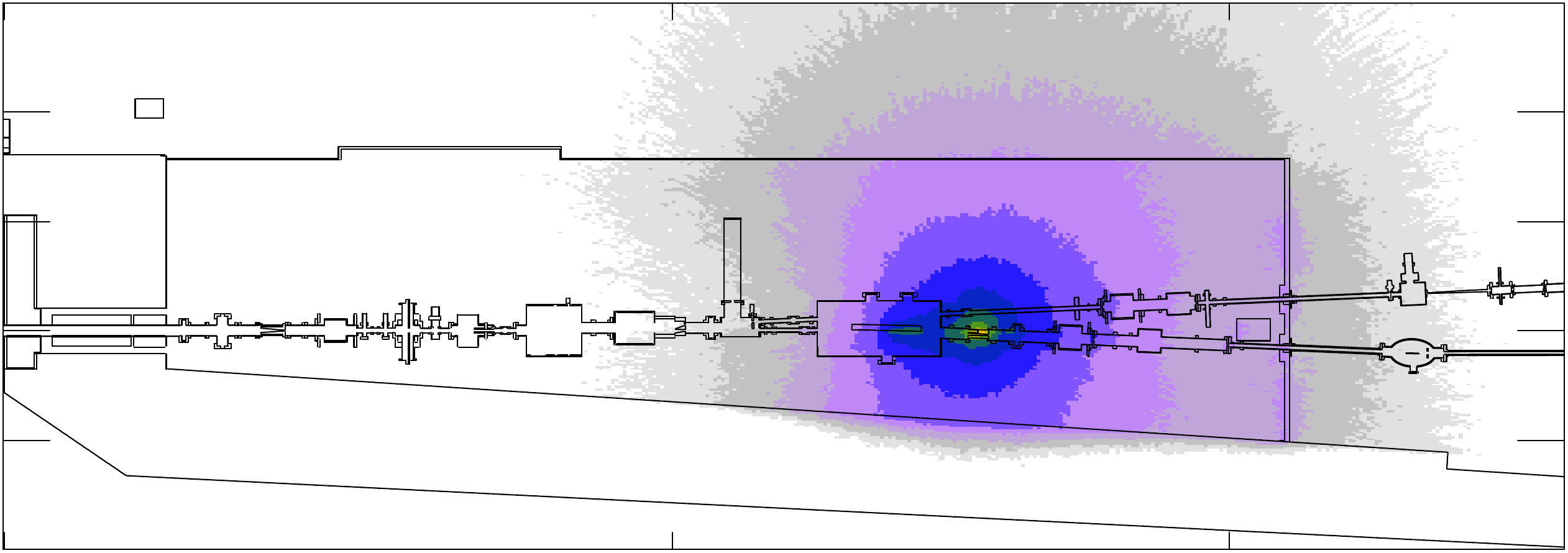}}  \\
\vspace{0.2cm}
\emph{photon H*} \\
\vspace{0.1cm}
\includegraphics[scale=0.182]{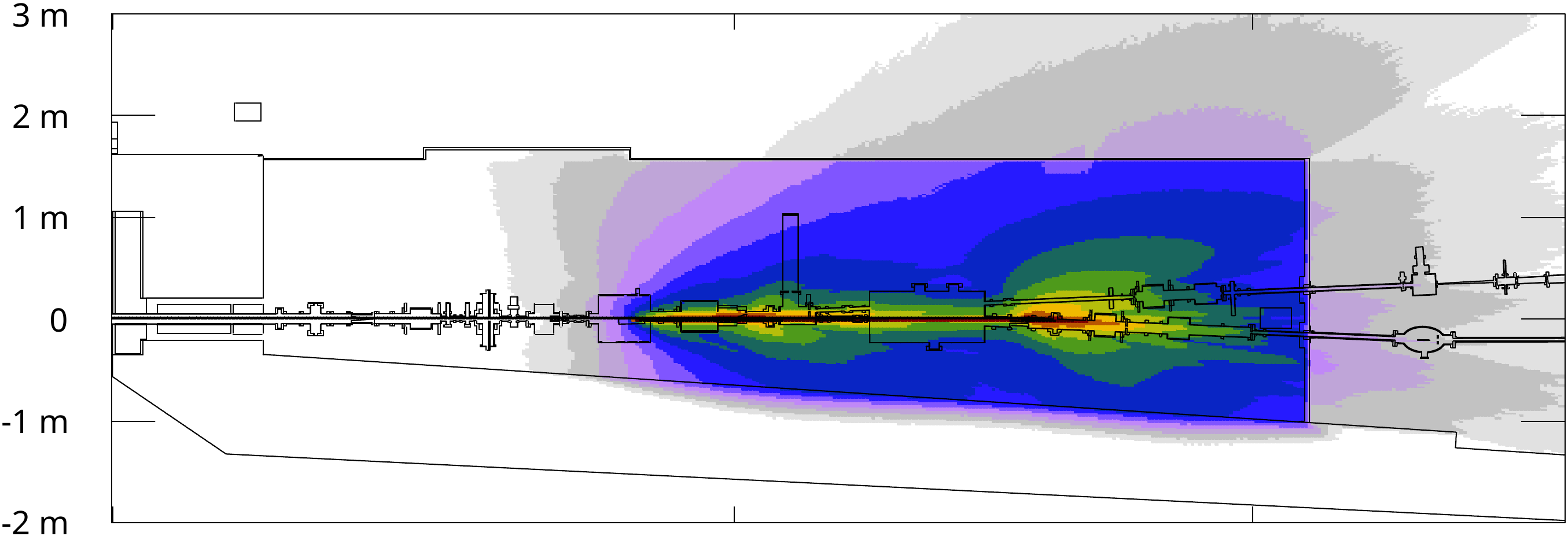}\hspace{0.2cm}\raisebox{0.6mm}{\includegraphics[scale=0.182]{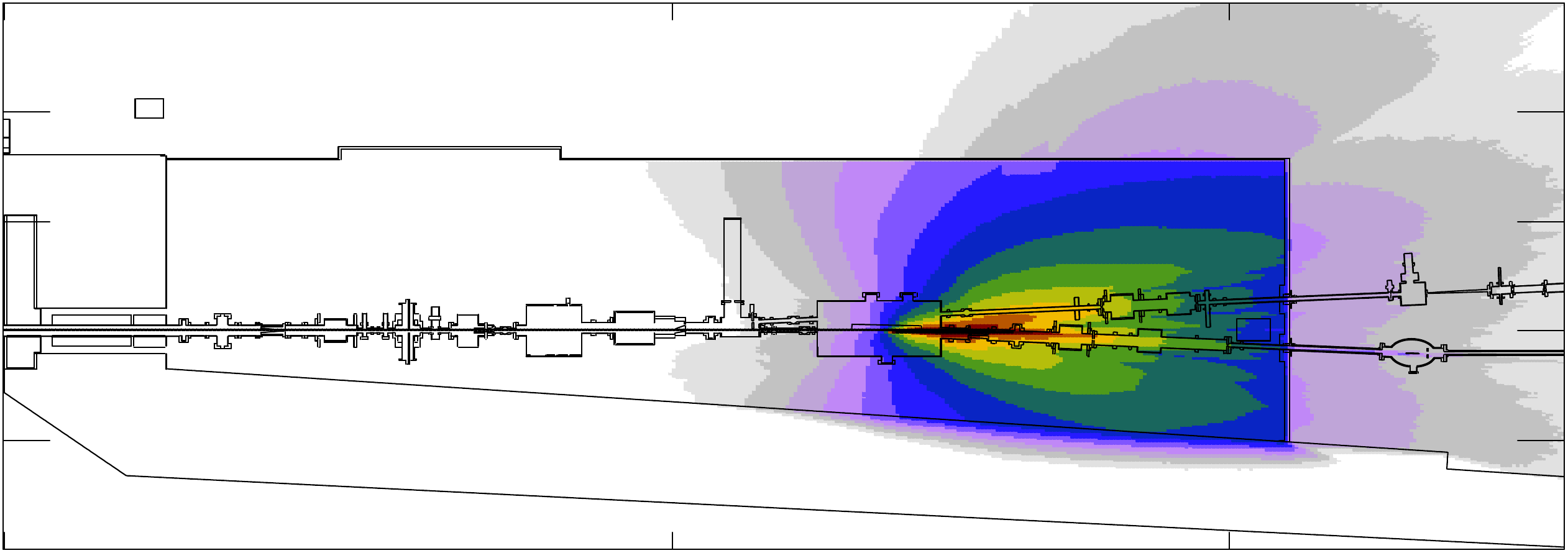}}  \\
\vspace{0.2cm}

\includegraphics[scale=0.4]{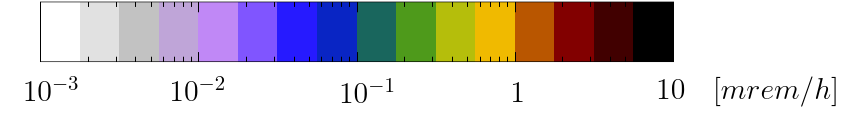}  \\
\vspace{0.1cm}
\caption{\it  \label{GBD} NSLS-IIU 3, case A (left) and NSLS-IIU 3.5, case B (right) respective neutron and photon H* rates [\si{\milli rem \per\hour}] resulting from GB impinging onto scattering mirrors at full power. The results are plan views averaged over $Y=\pm1$ \si{\meter}, centered at the beamline elevation ($1$ \si{\milli rem} $\equiv 10$ \si{\micro\sievert}).}
\end{figure*}

 In figure \ref{GBD}, the distribution of GB-induced H* rates for NSLS-IIU 3 (A) and NSLS-IIU 3.5 (B) due to photon and neutron radiation are shown.

The overall H* and H*(10) results outside of the steel FOE variant due to GB (case A) are shown in table \ref{tab:sys4}. Only results on contact with the outboard walls and roof are listed because the downstream wall remained unchanged, all values are within a $10\%$  error range.

\begin{table}[htbp]
\caption{\label{tab:sys4} Dosimetry analysis on contact outside of the FOE steel outboard wall and roof for GB (case A) at full power. Values are H* | H*(10) [\si{\micro\sievert\per\hour}].}
\centering
\begin{tabular}{lcc}
 & Outboard wall  & Roof     \\

\hline

NSLS-II   &  \it \small .43 | .47 &  \it \small .50 | .44   \\
  
  NSLS-IIU 3   &  \it \small .52 | .56 &   \it \small .60  | .53 \\

  NSLS-IIU 3.5   &  \it \small .40 | .44 &  \it \small .51 | .49 \\

  NSLS-IIU 4   &  \it \small .36 | .35 &  \it \small .40 | .42   \\
\end{tabular}
\end{table}

\begin{figure}[htbp]
\centering

{\includegraphics[scale=0.1921]{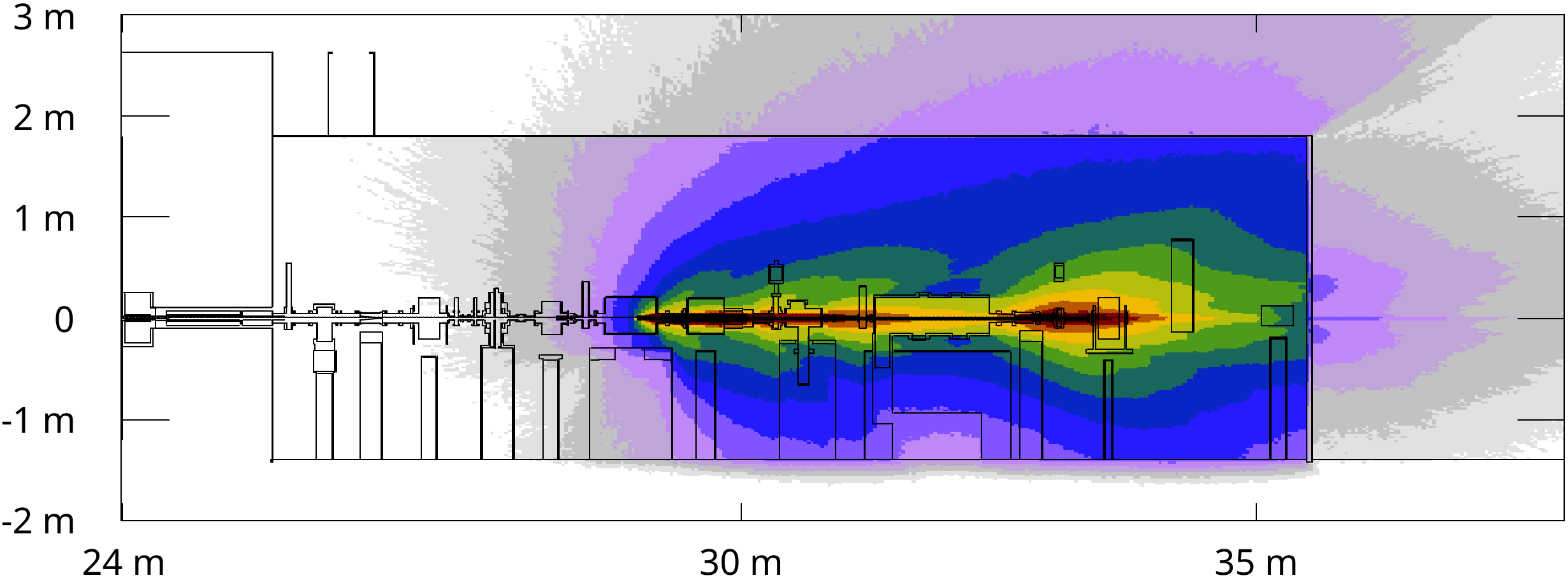}}  \\

\includegraphics[scale=0.1921]{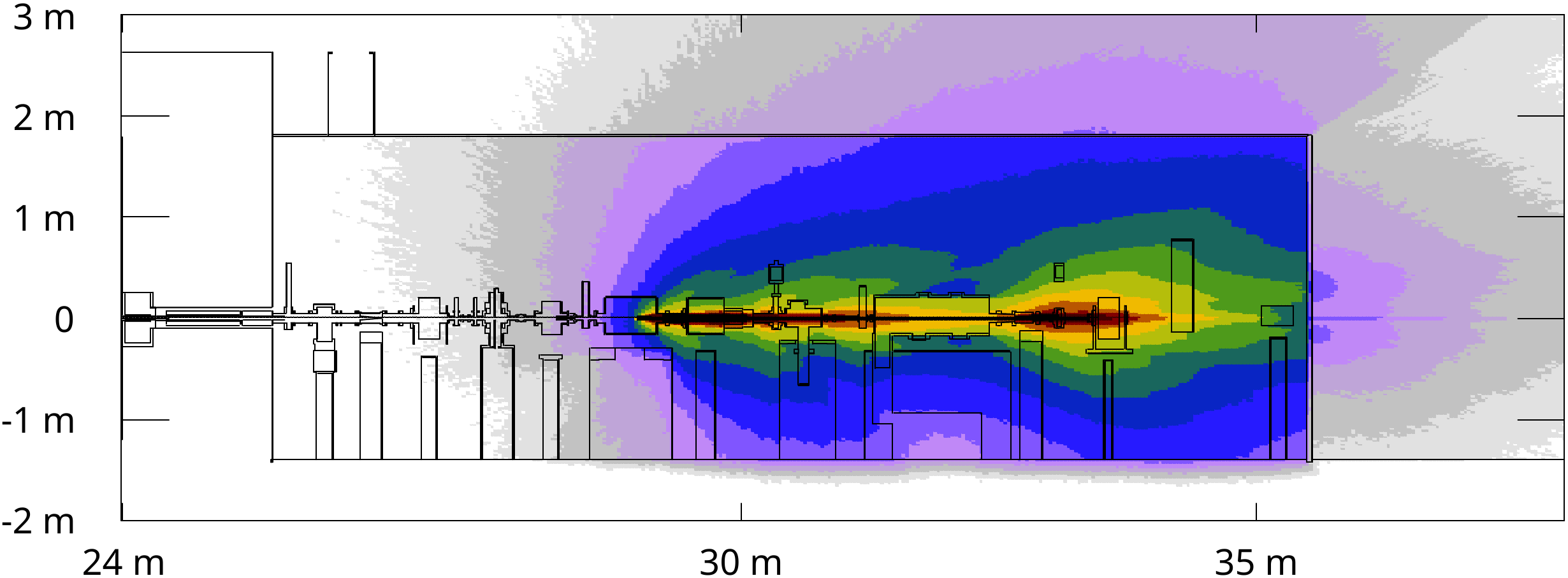}  \\

\vspace{0.3cm}

\includegraphics[scale=0.3]{scala.png}

\caption{\it  \label{GBI} NSLS-IIU 3.5 GB case A's H* rates [\si{\milli rem \per\hour}] for lead (above)  and steel (below) FOE shielding. Results are elevation views averaged over 2 \si{\meter}  ($1$ \si{\milli rem} $\equiv 10$ \si{\micro\sievert}).}
\end{figure}

The dosimetry results for NSLS-IIU 3.5 case A, with the FOE steel variant shielding compared against the standard lead FOE shielded version, are shown in figure  \ref{GBI}.

\begin{figure}[htbp]
\includegraphics[scale=.29]{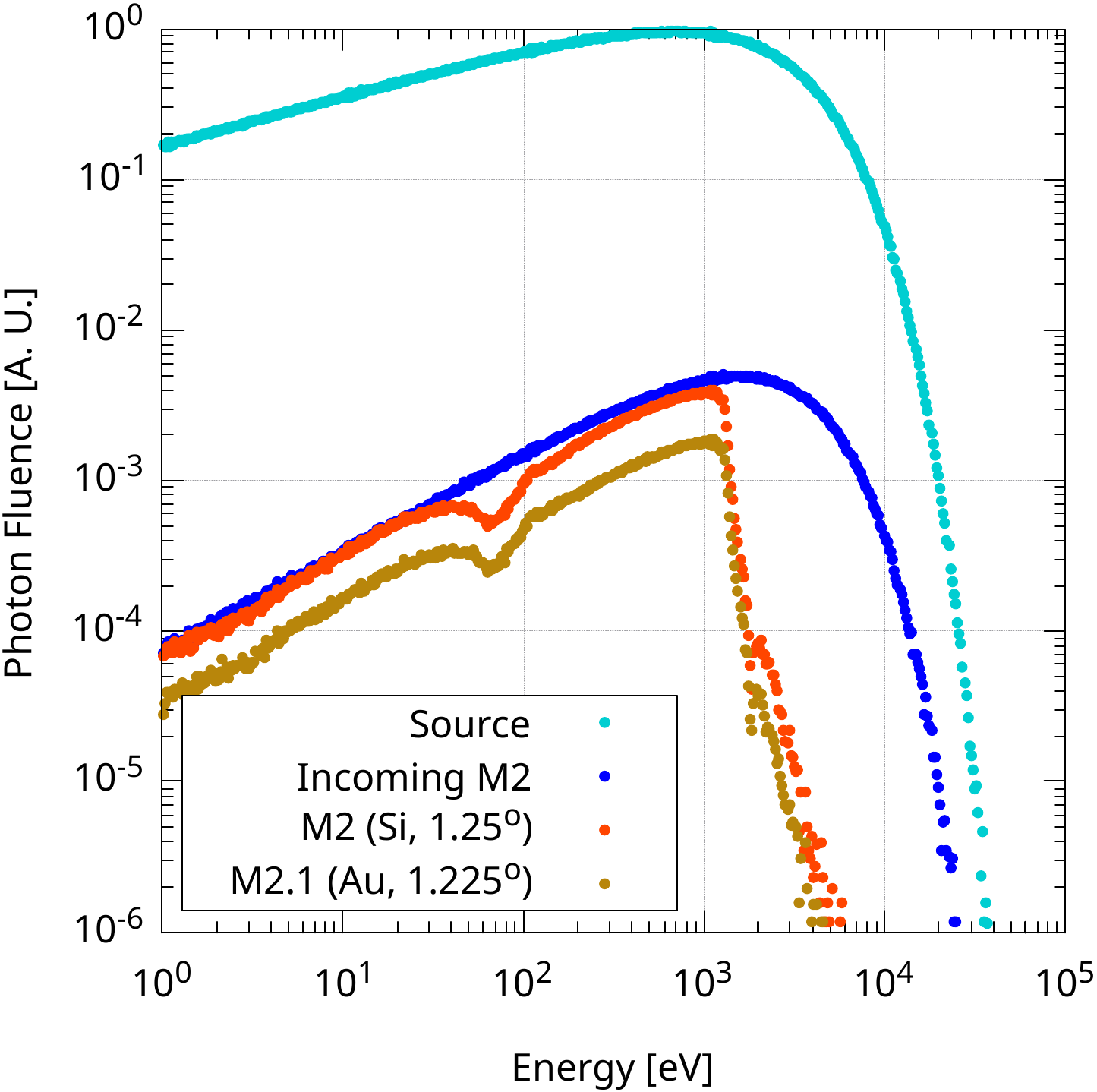}
\caption{\it \label{iso} Photon fluence (in arbitrary units) from the source and two consecutive reflections from a Si (1.25\si{\degree}) and then Au (1.225\si{\degree}) mirror.}
\end{figure}

\begin{figure}[htbp]
\centering
\includegraphics[scale=.515]{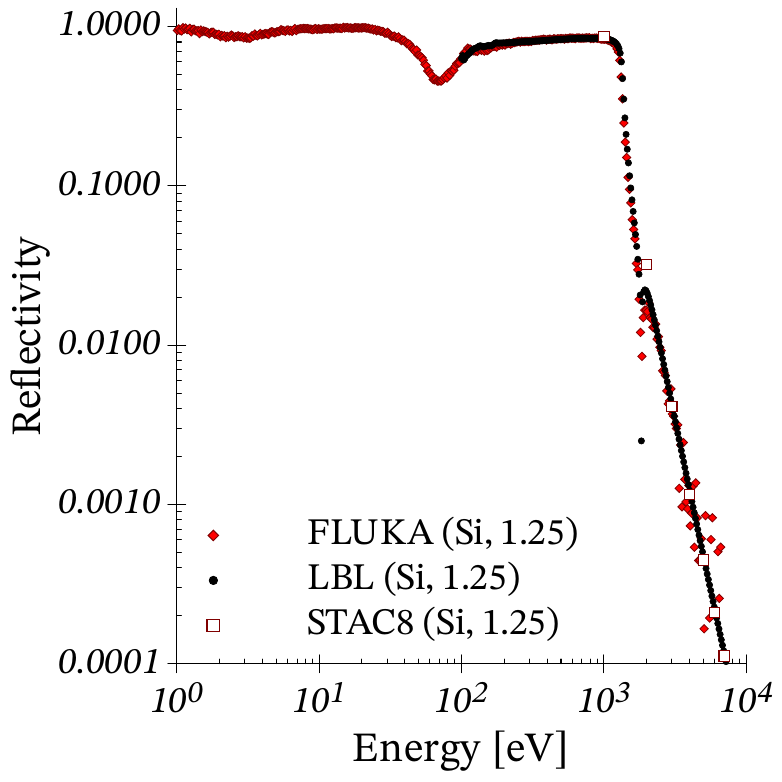}

\caption{\it  \label{reff} Reflectivity coefficients calculated with FLUKA for the M2 mirror  (Si, 1.25\si{\degree}), a comparison with LBL's X-ray database and with values calculated with STAC8 is also given.}
\end{figure}

An example of the reflection effect of each of the two mirrors in the inboard section is provided in figure \ref{iso}, as calculated with FLUKA. The reflectivity coefficients obtained with FLUKA are detailed in figure \ref{reff} for the M2 (Si) mirror. A benchmarking is also provided therein with the reflectivity coefficients from STAC8 and the X-ray reflectivity database from LBNL's CXRO \cite{lbl}.

\begin{figure}[htbp]
\centering
\includegraphics[scale=.19]{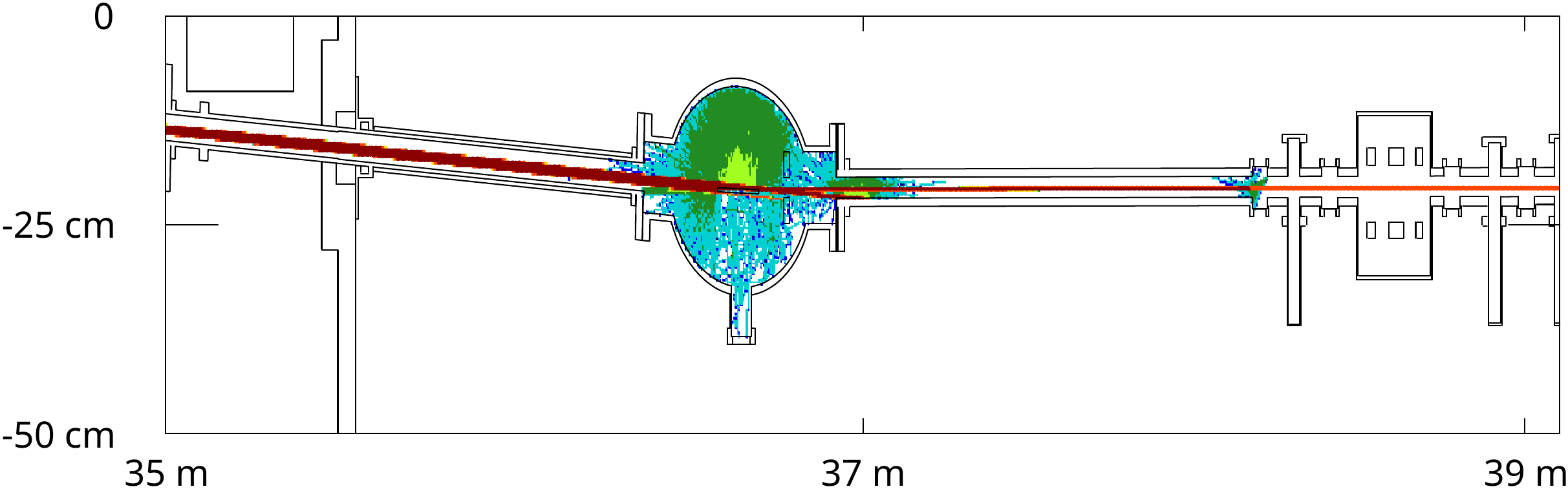}\\

\vspace{0.2cm}
\includegraphics[scale=.16]{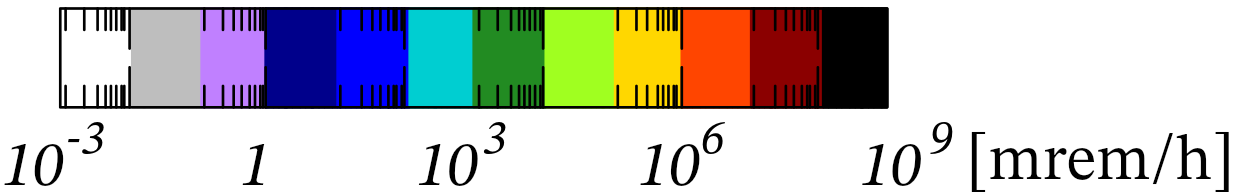}
\caption{\it  \label{dosesr} Radiation distribution in DIM's inboard branch, when the beam reflected by mirror M2.1 intercepts/grazes a Si conical scattering element just upstream of a thin steel bellow ($1$ \si{\milli rem} $\equiv 10$ \si{\micro\sievert}).
}
\end{figure}

\begin{figure*}[htbp]
\centering
 \hspace{3.5cm} \: \hspace{9.5cm}  \footnotesize [mrem h$^{-1}$]\hspace{3cm}\normalsize\\
\includegraphics[scale=.33]{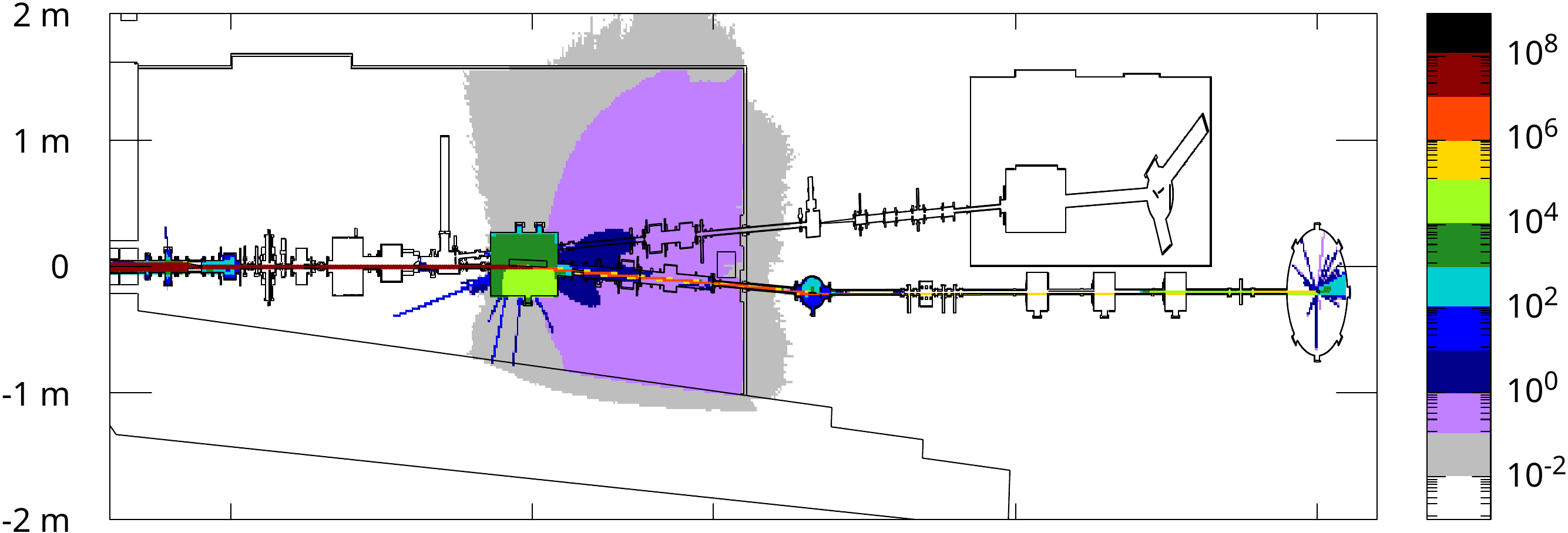}\\
\caption{\it  \label{faultz} Combined dosimetry analysis of nominal maximum GB and SR derived from fault scenarios at the outboard (top panel) and inboard (bottom panel) sections of the beamline. The results are given in mrem h$^{-1}$ and averaged over a 2 \si{\meter} elevation, centered at the straight section ($1$ \si{\milli rem} $\equiv 10$ \si{\micro\sievert}).}
\end{figure*}

The ambient dose distribution due to SR  propagating throughout the beamline, from M2 to M2.1, and then grazing the scatterer in the way towards the endstation, is detailed in figure  \ref{dosesr}. A combined SR fault scenario and GB full power dose map is shown in figure \ref{faultz}.

\section{Discussion}

The highest dose rates compiled in table \ref{tab:fantastictoss} are scored outside of the outboard of the ratchet wall for case 2, with radiation levels incrementing by $\sim 15\%$ at each upgrade stage.  This radiation leak develops at the weakest section of the storage ring shielding, immediately downstream of the sliding door. By placing the ARMs near that hotspot one effectively guarantees that a fault conditions is promptly detected and beam is suppressed within seconds, thus preventing any significant radiation exposure to personnel. In case 1 the magnitude of the maximal leaked radiation through this weak spot is about an order of magnitude lower than case 2 since the beam interacts further upstream with the FE slits and thence farther from the ratchet wall.

H* values scored in table \ref{tab:fantastictoss} were found to be systematically lower than H*(10). However, where the higher energy radiation (possibly close to 1 \si{\giga\electronvolt}) contributes significantly some differences can be observed. This is visible in figure \ref{comp}, in the forward projected cascade shape immediately after the beam interaction with the slit.
Conversely, H*(10) weight coefficients tend to be overall higher since the energies of particles are relatively low to see any increment from the H* conversion factors. On the other hand the electrons and positrons tend to have higher weight coefficients at the tens of \si{\mega\electronvolt} in the H*(10) case, causing the H*(10) plume inside the FOE to be much more intense than that of the H* case. These effects are also confirmed by the spectrum shown in figure \ref{blau}, scored at the ARM detector close to the main radiation leak in NSLS-IIU 4; case 2. In the latter, the weighted fluence for photons in H* only becomes larger than that of H*(10) at $\sim 15$ \si{\mega\electronvolt} whilst the weighted fluence for electrons and positrons below $\sim 20$ \si{\mega\electronvolt} is almost one order of magnitude higher for H*(10) with respect to H*.

From the GB dosimetry results in table \ref{tab:sys3}, NSLS-IIU 3 can be considered the worst case scenario. Note that the baseline results already exceeds the NSLS-II radiological goals as the secondary GB shielding is not optimized. The lower energy scenarios tend to be more conservative towards the outboard wall and roof of the FOE while the the higher energies cases could have a potentially higher impact on the downstream shielding if not for their lower intensity. For this reason, NSLS-IIU 4 is the scenario with the least impact from a radiation safety standpoint.

H*(10) rates tend to be higher for the results downstream of shielding in table \ref{tab:sys3}, by as much as $20\%$ in most cases. On the other hand, H* rates are only  higher for the outboard results at 3 \si{\giga\electronvolt} beam energies, locally driven by scattered electromagnetic radiation.

Figure \ref{GBD}'s dose maps help to illustrate the distribution of neutrons and photons in the FOE. Firstly, the neutronic component is isotropic and emanates essentially from  the main interaction points. Secondly, the electromagnetic component tends to be strongly forward projected, resulting in higher dose rates downstream of the FOE.

Regarding the usage of steel to shield the roof and outboard wall of the FOE, the GB results compiled in table \ref{tab:sys4} confirm that the proposed thicknesses of steel can achieve a superior shielding performance in the DIM beamline  with respect to the standard FOE results indicated in \ref{tab:sys3}. This observation is also confirmed by the virtually equivalent dose maps detailed in figure \ref{GBI}. Since lead is more effective than steel against electromagnetic radiation, it is necessary to keep the downstream wall unaltered in the FOE steel variant  because otherwise a prohibitively large thickness of steel would be required.

FLUKA's new capabilities in handling X-ray reflections allow for the transport of the generated radiation throughout a beamline's optical elements (see figure \ref{iso}). A comparison of the reflectivity coefficients obtained independently across three sources yielded a relatively good agreement as detailed in figure \ref{reff}. It should be noted that this threefold comparison is only completely valid for the 1-10 \si{\kilo\electronvolt}, since STAC8 and LBL data do not extend below 1 \si{\kilo\electronvolt} and 100  \si{\electronvolt}, respectively.

This new FLUKA feature can be also employed to extract radiological information from scenarios involving reflected X-rays and help validate shielding requirements. A demonstration is displayed in figure \ref{dosesr}, depicting the partial interaction of the reflected beam with a scatterer element downstream of the M2.1 mirror. The SR is transported throughout the beamline, from the bending magnet source to the endstation, within the same simulation and in a single step. 
 Moreover, a radiological analysis can be further ``strengthened'' by combining the SR and GB dosimetry output,  as denoted in figure \ref{faultz}. The advantage of the aforementioned methodology is to provide a more complete and accurate dosimetry result since those radiological source terms are the main contributors to radiation levels throughout NSLS-II beamlines.

\section*{Conclusion}

This study consisted of a preliminary shielding analysis using the FLUKA code and a relatively conservative beamline design, aimed at assessing the potential impact of three NSLS-II upgrade scenarios, with variations of the FOE shielding and updated radiological estimators for external radiation exposure. 

Assuming an equivalent injection rate for all cases, the injection fault studies showed that the NSLS-IIU 4 upgrade yielded the highest dose rates, particularly at an (expected) hotspot outside of the storage ring ratchet wall, which is monitored by an ARM.
 Elsewhere, the upgrades' radiological impact remained within a factor of two of the current baseline.

 The GB results were more stringent overall for the upgrade NSLS-IIU 3, as the intensity increase yielded the highest dose rates outside of the FOE compared to the  baseline results and the energy upgrades. Still comparing to the baseline, energy upgrades have the potential to create slightly higher radiation leakage downstream of the FOE, driven largely by electromagnetic radiation.

The steel variant considered for the FOE proved equally effective against the GB source terms corresponding to energy upgrades, and significantly more effective for both baseline energy cases. These results indicate that the steel variant proposed could indeed be regarded as a valid alternative for future FOE designs. 

The use of the H* estimator was found to yield generally lower values than the currently employed H*(10), this was attributed to the lower energy range of radiation exiting the shielding and the respective lower H* dose coefficients for electrons in particular. Thus, an eventual adoption of this radiological quantity would not have a detrimental effect on NSLS-II operations.

Finally, a demonstration of transport of SR from source to endstation, utilizing the newly developed X-ray reflectivity FLUKA capabilities was provided in this study. The demonstration was further extended to highlight the radiological consequences of a fault scenario, previously only evaluated with analytical codes. For completeness, the GB and SR contributions were also combined, providing a visualization of the total dose rates in a single result. The latter methodology fully exploits the predictive capabilities of the FLUKA code to assess the radiological impact of a particular beamline configuration and shielding design.

\section*{Acknowledgments}

The NSLS-II at BNL is supported by the U.~S. Department of Energy, Office of Science, Office of Basic Energy Sciences. The authors thank NSLS-II colleagues for the frequent discussions and the FLUKA Collaboration, particularly Dr. Alfredo Ferrari,  for the X-ray reflectivity incorporation.\\

\vfill

%
%
%

\end{document}